\documentclass[prb,preprint,amsmath,amssymb]{revtex4}
\usepackage{graphicx,dcolumn,bm,color} 

\begin{document} 

\title{Semimetal-to-Antiferromagnetic Insulator Transition in Graphene Induced by Biaxial Strain}
\author{Sung-Hoon~Lee} \email{sung-hoon.lee@samsung.com}
\affiliation{Computational Science Group, Samsung Advanced Institute of Technology, Yongin 446-712, Korea}
\author{Sungjin~Kim} 
\author{Kinam~Kim} 
\affiliation{Computational Science Group, Samsung Advanced Institute of Technology, Yongin 446-712, Korea}
\date{\today}

\begin{abstract}
{ 

We report first-principles calculations on the antiferromagnetic spin ordering in graphene under biaxial strain. 
Using hybrid functional calculations, we found that the semimetallic graphene sheets undergo a transition to antiferromagnetic insulators at a biaxial strain of 7.7\% and that the band gap rapidly increases after the onset of this transition before reaching 0.9 eV at a biaxial strain of 12\%. 
We examined the competition of the antiferromagnetic spin ordering with two-dimensional Peierls distortions upon biaxial strain, and found that the preceding antiferromagnetic insulator phase impedes the Peierls insulator phase. 
The antiferromagnetic insulator phase is destabilized upon carrier filling but robust up to moderate carrier densities.
This work indicates that biaxially strained graphene represents a noble system where the electron-electron and electron-lattice interactions compete with each other in a simple but nontrivial way.

}
\end{abstract}

\maketitle \newpage

\section{Introduction}

Graphene, which consists of a honeycomb lattice of carbon atoms, exhibits a high carrier mobility and is appealing for electronic applications, such as channel materials for field-effect transistors.\cite{Novoselov05,Zhang05} A crucial obstacle for the potential applications of graphene is its absence of a band gap, which results in poor on--off current ratios in a transistor.\cite{Schwierz10} The introduction of a band gap or a metal-to-insulator transition in semimetallic graphene can greatly enhance its utility.

Graphene is a semimetal with an eight-fold degeneracy at its intrinsic Fermi level, which originates from three twofold symmetries of sublattices, valleys, and spins.\cite{CastroNeto09} Therefore, opening the band gap requires breakage of the symmetries or mixing of the degenerate states.  The gap opening in graphene on SiC substrates \cite{Zhou07} is induced by sublattice symmetry breaking,\cite{Semenoff84} which results in an asymmetric charge population in two sublattices of the honeycomb lattice. The degeneracy from two inequivalent Dirac valleys at $K$ and $K'$ can be lifted by intervalley mixing potentials with a real-space periodicity of $(\sqrt{3}\times\sqrt{3})R30^\circ$.\cite{Manes07,Chamon00} The gap opening in graphene with periodic adsorbates \cite{Balog10} or holes \cite{Pedersen08,Kim10} and in armchair-edge graphene nanoribbons \cite{Barone06,Son06b,Han07} belongs to this category.\cite{Lee11} Recently, we have shown \cite{Lee11} that gap opening by intervalley mixing occurs in biaxially strained graphene through spontaneous inverse-Kekul\'e distortions that produce potential modulations of the $(\sqrt{3}\times\sqrt{3})R30^\circ$ periodicity rather than through external modulations. However, this two-dimensional Peierls metal--insulator transition is accompanied by structural failure.\cite{Lee11,Marianetti10} The physical origin of the gap opening under biaxial strain is distinct from that under uniaxial strain \cite{Pereira09a,Pereira09b,Choi10} or inhomogeneous strain,\cite{Guinea09} which breaks local hexagonal symmetries to induce Dirac-point merging or pseudo-Landau level formation, respectively.  As the last category, the gap opening in zigzag-edge graphene nanoribbons involves breaking of the spin symmetry by the formation of ferrimagnetic spin ordering at the edges.\cite{Fujita96,Son06b} 

In this work, we investigate the possibility of spin ordering in graphene, not just at edges, but throughout the two-dimensional sheet. The honeycomb lattice, which is composed of two triangular sublattices, can support unfrustrated antiferromagnetism and can have an antiferromagnetic ground state with a finite band gap when the onsite electron--electron Coulomb energy, $U$, is high compared to the electronic hopping integral, $t$.\cite{Sorella92} The critical value of $U/t$ for the antiferromagnetic spin ordering has been predicted to be $3.6 \sim 4.3$ according to quantum Monte Carlo calculations based on the Hubbard model.\cite{Sorella92,Martelo97,Furukawa01,Meng10} In Fig.~1, the electronic ordering of the antiferromagnetic insulator phase is compared to those of the other gap-opening orderings. The antiferromagnetic insulator phase also breaks the sublattice symmetry; however, in this case, the symmetry breaking involves spin ordering, not charge ordering. Graphene, whose electronic structure is characterized by the honeycomb lattice of C $2p_z$ states, has a zero band gap with no magnetic ordering because the relative onsite Coulomb energy $U/t$ is subcritical.\cite{Yazyev10,Wehling11} The ferrimagnetic ordering at the zigzag edges can be attributed to a local increase in the effective $U/t$ at the edges by reduced bonding configurations. A homogeneous increase in $U/t$ can be achieved if the atomic distances are uniformly increased, e.g., through the application of biaxial strain. 

In this study, we used first-principles calculations to examine the onset and stability of the antiferromagnetic insulator phase in biaxially strained graphene.  Our hybrid functional calculations showed that the semimetal-to-antiferromagnetic insulator transition occurs at a biaxial strain of 7.7\% and that the band gap rapidly increases after the onset of this transition before reaching 0.9 eV at a biaxial strain of 12\%. We also examined the competition of the antiferromagnetic spin ordering with two-dimensional Peierls distortions upon biaxial strain, and found that the preceding antiferromagnetic insulator phase impedes the Peierls insulator phase.  Our calculations on the effect of carrier filling showed that the antiferromagnetic insulator phase is destabilized upon carrier filling but robust up to moderate carrier densities.

\section{Method}

To study the antiferromagnetic spin ordering, we performed hybrid functional calculations \cite{Heyd03} in which the exact Hartree--Fock (HF) exchange energy was hybridized with the exchange-correlation energy from the generalized gradient approximation (GGA) within the framework of a generalized Kohn--Sham scheme \cite{Seidl96} to remedy the self-interaction error of the GGA calculations.\cite{Kudin02}  Hybrid functional calculations have been used to describe the electronic structures of traditional $3d$ and $5f$ antiferromagnetic Mott insulators, such as NiO, MnO, VO$_2$, and UO$_2$.\cite{Kudin02,Marsman08,Archer11,Eyert11,Crespo12}

Our hybrid functional calculations employ the hybrid functional of Heyd, Scuseria, and Ernzerhof \cite{Heyd03} and the projector-augmented-wave method, as implemented in VASP.\cite{Kresse96,Kresse99} Valence electronic wavefunctions are expanded in a plane-wave basis set with a cutoff energy of 400 eV. In our supercell calculations, the graphene layers are separated by 10 \AA. The k-point integration was performed at a uniform k-point mesh of ($30 \times 30$) in the Brillouin zone of the $(\sqrt{3}\times\sqrt{3})R30^\circ$ cell. The atomic positions are relaxed until the residual forces become less than 0.001 eV/\AA. Our methods predict the band gaps of diamonds and hexagonal BN sheets to be 4.8 eV and 4.5 eV, respectively, compared to the experimentally obtained values of 5.5 eV and 5.2 eV, respectively.

\section{Results and Discussion}

We examined the semimetal-to-antiferromagnetic insulator transition in graphene by increasing the lattice constant from a calculated equilibrium value. The antiferromagnetic phase becomes stable over the nonmagnetic phase at the relative increase in the lattice constant, or at an equibiaxial strain, $\varepsilon$, of $\sim 7.7$\% (Fig.~2a).  The band structure after the transition (Fig.~2b) exhibits a gap at the $K$ point. Here, each of the energy bands is doubly degenerate, which corresponds to spin-up and spin-down components. The electronic wavefunctions of the two spin components (Fig.~2c) reveal contrasting spatial populations. The electronic states of the valence bands exhibit a preferential occupation of the spin-up component at the A sublattice and that of the spin-down component at the B sublattice, which accounts for the antiferromagnetic spin ordering. For the conduction bands, the spatial occupation is reversed for the two spin components. The calculated band gap (Fig.~2d) exhibits an almost identical dependence on $\varepsilon$ with the staggered magnetic moment (Fig.~2a), which clearly reveals the correlation between the spin ordering and the band-gap opening. 

The transition to the antiferromagnetic insulator phase can be attributed to the reduction of the hopping integral, $t$.  The early onset of the antiferromagnetic insulator phase at $\varepsilon = 7.7$\% in graphene is remarkable because the typically accepted value of $U/t$ for pristine graphene, $\sim 1.2$,\cite{Yazyev10} is considerably smaller than  $(U/t)_c = 3.6 \sim 4.3$, and the hopping integral, $t$, only decreases by $\sim 15$\% at $\varepsilon = 10$\% (See Fig.~3).  However, a recent evaluation \cite{Wehling11} of the Coulomb interaction through a combination of first-principles calculations and a many-body formulation revealed that the $U/t$ of pristine graphene is as large as 3.3 and increases to $\sim3.8$ at $\varepsilon = 10$\%. The present hybrid functional calculations support the large value of $U/t$.

Because biaxially strained graphene can also exhibit a band gap opening as a spontaneous two-dimensional Peierls transition,\cite{Lee11} we studied the competition of the antiferromagnetic insulator phase with the Peierls insulator phase.
Peierls dimerization always occurs in one-dimensional atomic chains, irrespective of the force constant of atomic bonds.\cite{Peierls} However, in two-dimensional lattices, such as the honeycomb lattice, this dimerization can only occur if the force constant is relatively small compared to the electronic hopping integral, $t$.\cite{Mintmire92} In pristine graphene, strong planar $\sigma$ bonds hinder the Peierls distortion. However, biaxial strain rapidly weakens the $\sigma$ bonds and results in the Kekul\'e or inverse-Kekul\'e distortion,\cite{Lee11} which has been recently shown to be the unique Peierls dimerization mode for graphene sheets in the thermodynamic limit.\cite{Frank11}

Our nonmagnetic calculations reveal an onset of the spontaneous inverse-Kekul\'e distortion at $\varepsilon = 12.3$\% (Fig.~4a). Above this onset, the band gap rapidly increases to a maximum of 3.0 eV at $\varepsilon \approx 18$\% (Fig.~4b), where the $\sigma^*$ band begins to decrease below the empty $\pi$ band. Given the use of hybrid functional calculations in the present work, this result is consistent with previous GGA calculations,\cite{Lee11} where the critical $\varepsilon$ is 14.2\% with the maximum band gap of 2.1 eV. The structural failure induced by this Peierls ordering, as predicted in previous works,\cite{Lee11,Marianetti10} occurs at $\varepsilon = 14.9\%$ (See Fig.~5). Compared to the antiferromagnetic insulator phase that initiates at $\varepsilon = 7.7$\%, this Peierls insulator phase exhibits a delayed onset but also exhibit a faster increase in the energy gain and band gap (Figs.~4a\&b).

To understand the competition between the antiferromagnetic and Peierls insulator phases, we calculate energy profiles as a function of lattice distortion with the magnetic ordering on and off. At $\varepsilon = 14.5$\%, nonmagnetic calculations reveal a double-well energy profile (Fig.~4c) with two local minima that correspond to the stable structures of the inverse-Kekul\'e and Kekul\'e distortions, respectively, and a central maximum that corresponds to the undistorted structure. The inverse-Kekul\'e distortion (A, in Fig.~4c), where a six-membered ring structure similar to benzene is formed, is more stable than the Kekul\'e distortion (B, in Fig.~4c), which is characterized by ordered carbon dimers. When $\varepsilon$ is increased, both energy minima deepen, and the stability of the inverse-Kekul\'e structure increases over the Kekul\'e structure.

When the antiferromagnetic ordering is turned on, the double-well energy profile changes to a single-well energy profile through a significant energy lowering of the undistorted structure (Fig.~4c). The calculated staggered magnetic moment (Fig.~4d) indicates that the antiferromagnetic ordering is maximal in the undistorted structure and is suppressed near the local minima of the Peierls ordering, which reflects the different symmetries of the two orderings. When $\varepsilon$ is increased, the single-well energy profile becomes flatter, and, at the crossing point where $\varepsilon = 15.3$\%, the minimum moves continuously from the antiferromagnetic undistorted structure to the nonmagnetic inverse-Kekul\'e structure.  Therefore, the single-well energy profile for $\varepsilon < 15.3$\% indicates that the antiferromagnetic insulator phase impedes the manifestation of the Peierls ordering and thereby slightly improves the mechanical stability of graphene under biaxial strain.

The stability of the antiferromagnetic insulator phase for $7.7 < \varepsilon < 15.3\%$ assumes a half-filling of the $\pi$ band. The addition of charge carriers by filling the conduction bands or emptying the valence bands reduces the spin asymmetry at each atomic site, which reduces the band gap. For the potential use of biaxially strained graphene for field-effect transistors,\cite{Schwierz10} the behavior of this filling-control metal--insulator transition\cite{Imada98} is important because graphene has nonzero minimum carrier densities of 10$^{10}$--10$^{12}$ cm$^{-2}$ because of electron or hole puddles formed by charged defects or ripples.\cite{Martin08,Adam07} The calculated band gap with variable carrier filling (Fig.~6a) exhibits a maximum at the half-filling condition and a bell-shaped decrease upon the deviation from the maximum. The initial decrease of the band gap upon deviation from the half-filling condition reduces as $\varepsilon$ increases. Therefore, at $\varepsilon = 10\%$, the carrier filling of 10$^{12}$ cm$^{-2}$ decreases the band gap by $\sim 0.04$ eV, which indicates that the antiferromagnetic insulator phase is quite stable upon carrier filling of practical situations. Figure~6b summarizes the antiferromagnetic metal--insulator transition in graphene as a function of equibiaxial strain and carrier filling.

Finally we remark on the accuracy of our calculations in predicting the antiferromagnetic order. In the hybrid functional of Heyd, Scuseria, and Ernzerhof \cite{Heyd03} used in this work, the exchange energy is described by a 25\% HF exchange energy in combination with a 75\% GGA exchange energy. Calculations using such a mixing parameter reasonably predict structural properties and band gaps of many inorganic materials\cite{Marsman08}, but show a tendency to overestimate the antiferromagnetic coupling.\cite{Crespo12} On the other hand, for organic materials and graphene nanostructures, they underestimate band gaps and incorrectly predict carrier localization.\cite{Jain11,Sai11} Reasonable band-gap values were empirically obtained for organic systems by taking the mixing parameter as the inverse of the dielectric constant of the material.\cite{Marques11} The dielectric constant of graphene, which is $2.5\sim 4$, as determined from many-body perturbation calculations\cite{Wehling11,Schilfgaarde11} suggests that the proper HF contribution for graphene could be as high as 40\%, enhancing the antiferromagnetic coupling. Further studies are necessary to clarify this issue.

\section{Conclusion}

We have shown, using hybrid functional calculations, that biaxial strain can introduce a semimetal-to-antiferromagnetic insulator transition in two-dimensional graphene sheets. The transition occurs before the two-dimensional Peierls transition and thereby secures a margin before structural failure. This transition is robust upon finite carrier filling of practical situations.
This work indicates that the application of biaxial strain provides a useful means to manipulate the electronic properties of graphene and could be exploited for electronic and spintronic device applications. From a fundamental perspective, our analysis of the competition between the antiferromagnetic ordering and the Peierls ordering in graphene provides an understanding of how the competition in an elemental linear atomic chain extends in two dimensions.

\begin{acknowledgements}
The authors thank Ji~Hoon~Shim and Jongseob Kim for helpful discussions. 
\end{acknowledgements}


\newpage

\begin{figure}[h] \includegraphics[width=11cm]{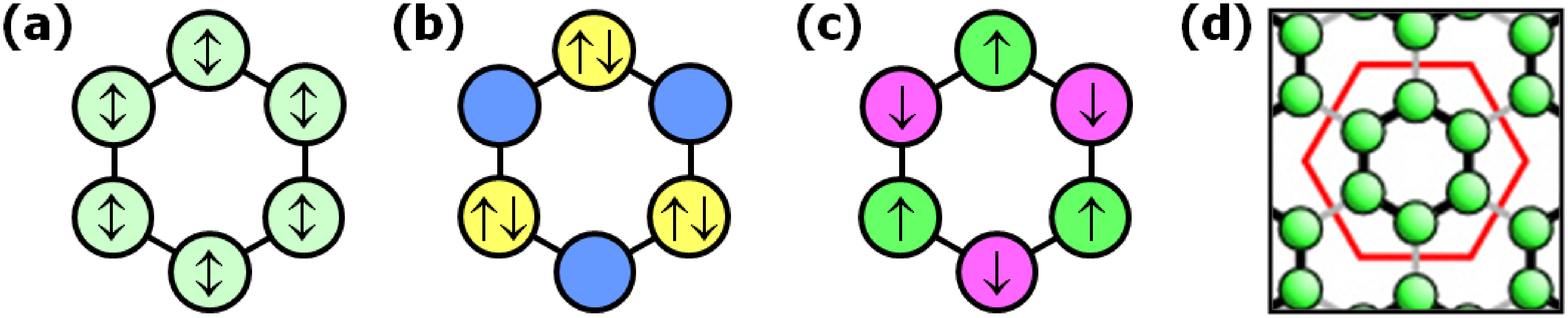} \caption{ (Color online) Band gap opening in graphene by charge, spin, and bond ordering. {(a)} The charge and spin population of pristine graphene; each C $2p_z$ orbital has a single electron without spin asymmetry. (b) The charge-ordering-induced sublattice symmetry breaking of hexagonal BN sheets. (c) The spin-ordering-induced sublattice symmetry breaking of the antiferromagnetic insulator phase. (d) The bond ordering of the Peierls insulator phase (the inverse-Kekul\'e distortion),\cite{Lee11} where the bond lengths of intracell and intercell bonds with respect to the Wigner-Seitz cell (red hexagon) differ. The inverse-Kekul\'e distortion produces intervalley mixing potentials. } \end{figure}

\begin{figure*}[h] \includegraphics[scale=1.2]{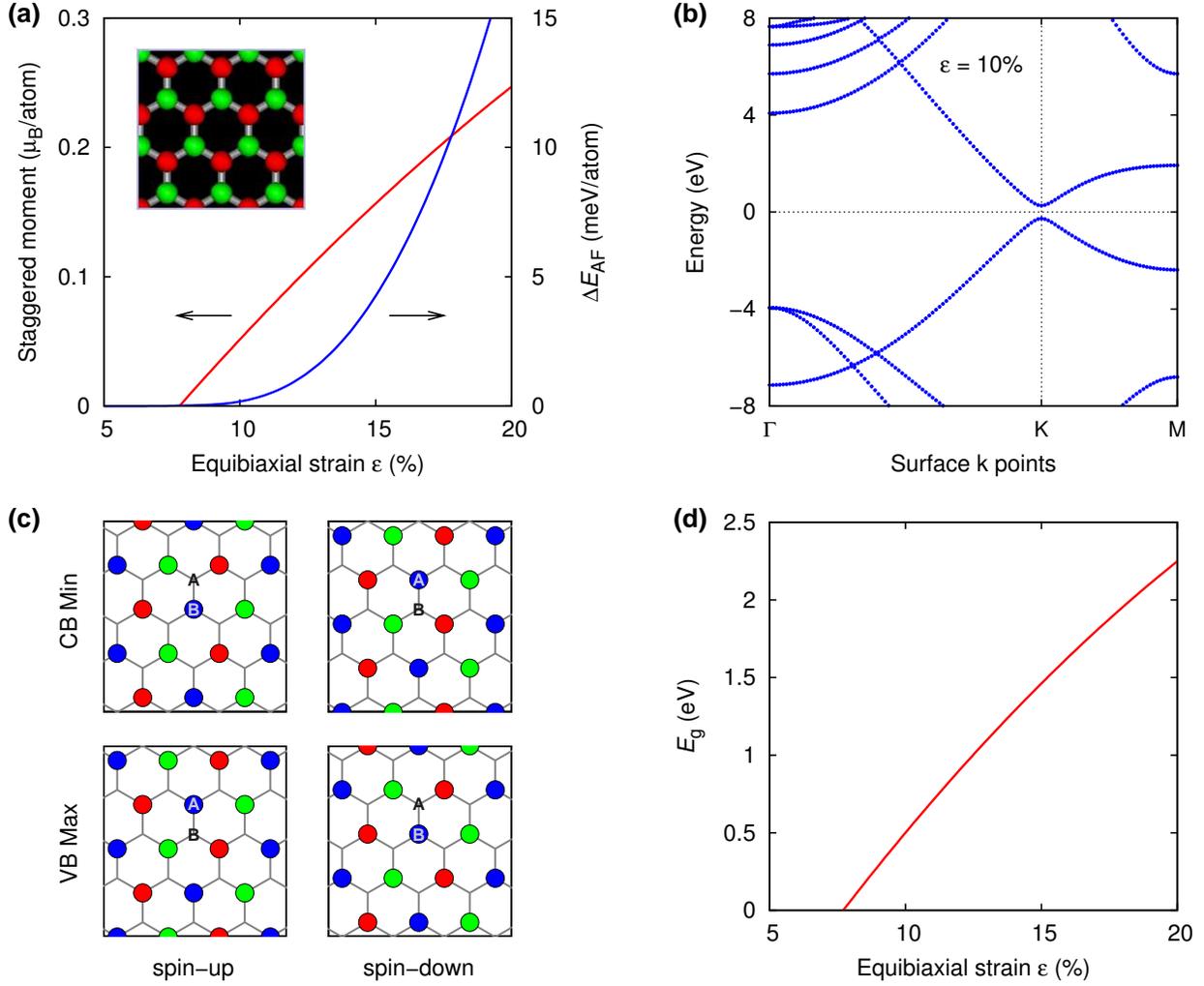} \caption{ (Color online) Semimetal to antiferromagnetic insulator transition in biaxially strained graphene. (a) The energy gain by antiferromagnetic ordering, $\Delta E_{\rm AF}$, and the staggered magnetic moment as a function of equibiaxial strain, $\varepsilon$. Inset, the spin density plot at $\varepsilon = 10\%$.  (b) The electronic band structure at $\varepsilon = 10\%$. In this figure, every band is doubly degenerate, which corresponds to spin-up and spin-down components. (c) The wavefunctions of the valence band maximum and the conduction band minimum for each spin component. The radius and color of a circle at each atomic site reflect the wavefunction coefficients of the C $2p_z$ orbitals with the phase angle of 0, $2\pi/3$, and $4\pi/3$, which are represented by red, green, and blue, respectively. (d) The band gap as a function of $\varepsilon$.} \end{figure*}

\begin{figure*}[h] \includegraphics[scale=1.2]{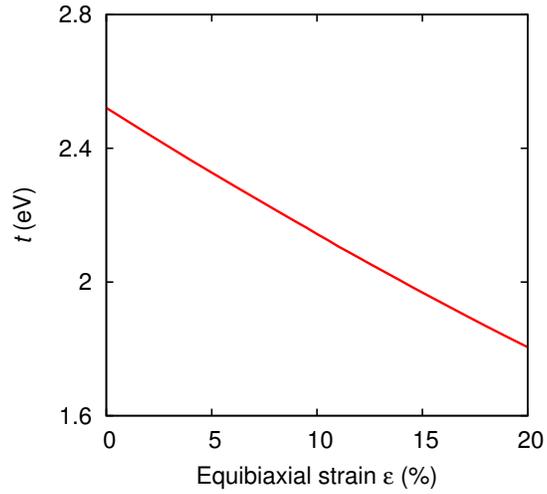} \caption{ (Color online) Hopping integral as a function of $\varepsilon$. The hopping integral between the nearest-neighbor sites, $t$, is estimated from the band gap at the $M$ point, which amounts to $2t$ in the nearest-neighbor tight-binding model\cite{CastroNeto09}. For semimetallic graphene, $t$ decreases almost linearly as $\varepsilon$ increases, i.e., $t=t_0(1-\alpha \varepsilon)$ with $t_0 \approx 2.5$ eV and $\alpha \approx 1.5$.
} \end{figure*}

\begin{figure*}[h] \includegraphics[scale=1.2]{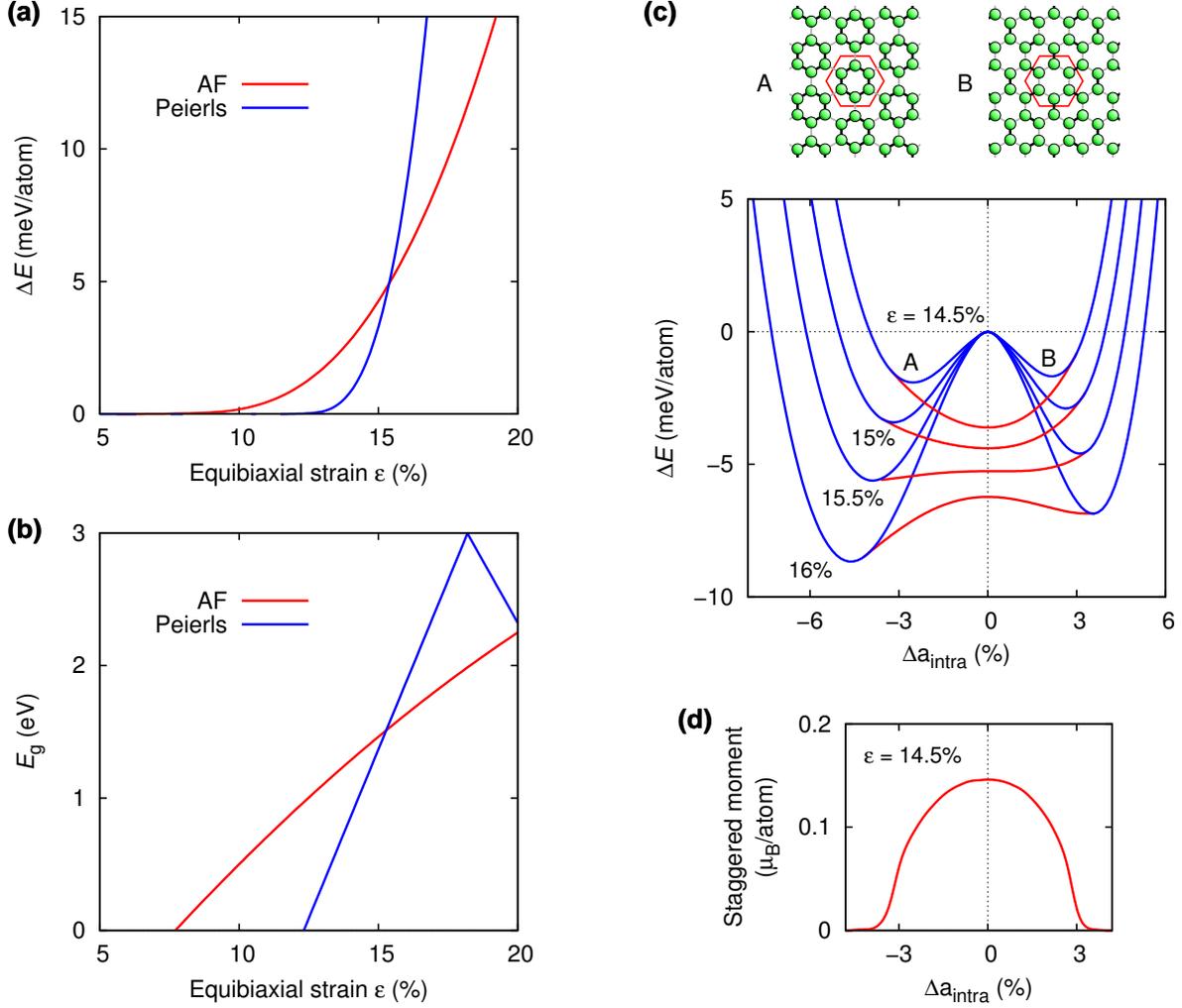} \caption{ (Color online) Antiferromagnetic (AF) versus Peierls insulator phases in biaxially strained graphene. (a) The energy gain of the antiferromagnetic and Peierls insulator phases over the semimetal phase. (b) The band gap of each phase; both of the energy gain and the band gap exhibit a crossover at $\varepsilon = 15.3\%$. (c) The energy gain of the Peierls insulator phase as a function of the relative intracell bond lengths; blue lines were obtained by spin-unpolarized calculations, and the red lines were obtained by spin-polarized calculations with antiferromagnetic spin ordering. The decrease and increase in the intracell bond lengths correspond to the inverse-Kekul\'e (A) and Kekul\'e (B) distortions (upper panels), respectively. (d) The staggered magnetic moment of the antiferromagnetic insulator phase as a function of the lattice distortion. } \end{figure*}

\begin{figure*}[h] \includegraphics[scale=1.2]{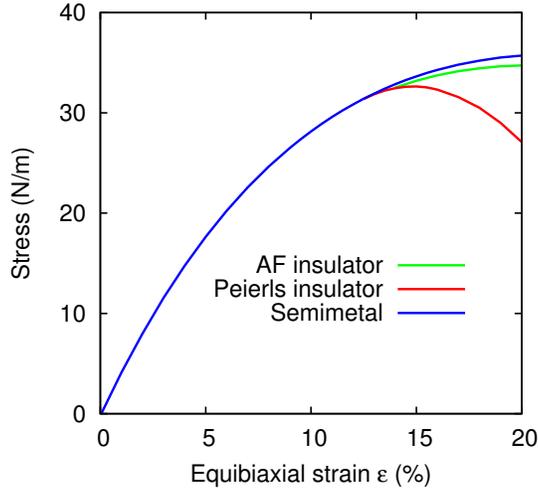} \caption{ (Color online) Stress as a function of biaxial strain $\varepsilon$. Whereas the antiferromagnetic (AF) spin ordering only slightly changes the mechanical properties, the Peierls ordering (the inverse-Kekul\'e distortion) decreases the mechanical instability point to $\varepsilon \approx 14.9$\%.
} \end{figure*}

\begin{figure*}[h] \includegraphics[scale=1.2]{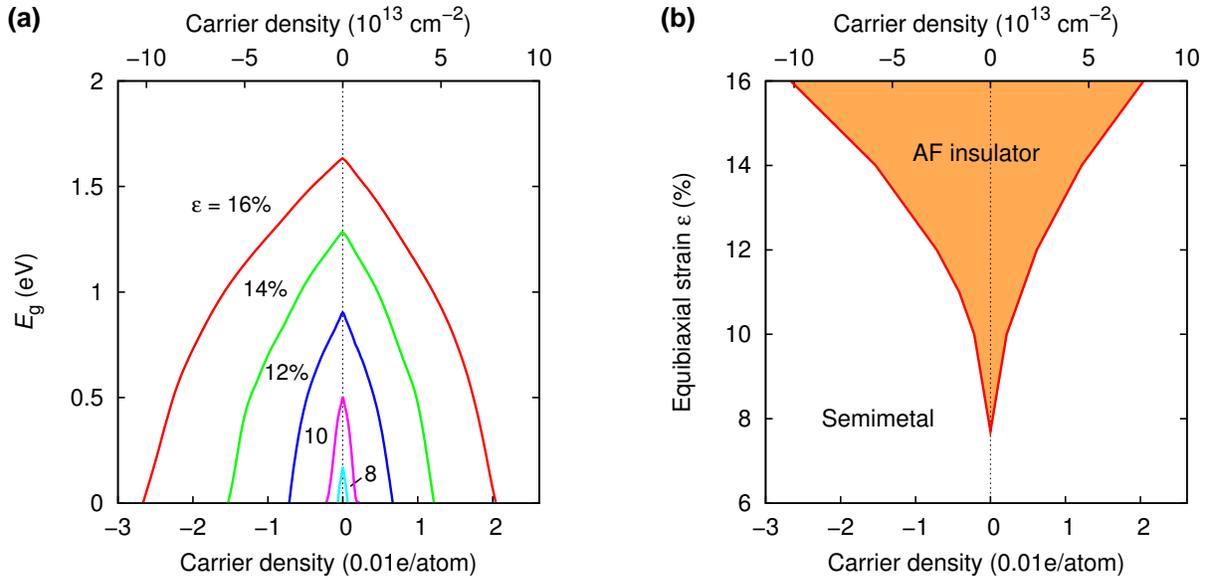} \caption{ (Color online) Filling-control metal--insulator transition of the antiferromagnetic insulator phase. (a) The band gap of the antiferromagnetic insulator phase as a function of carrier filling for $\varepsilon =$ 8--16\%. (b) The metal--insulator phase diagram in the plane of carrier density and biaxial strain, which was obtained from (a). Lattice distortions are not considered here.  } \end{figure*}

\end{document}